\documentclass[iop]{emulateapj}

\usepackage{ulem}
\usepackage{color}
\usepackage{soul}
\usepackage{amsmath}
\newcommand{\Msun}{\ensuremath{M_{\odot}}}

\usepackage{textcomp}
\usepackage{amssymb}
\shorttitle{Multiwavelength Observations of GX 339-4}
\shortauthors{Din\c{c}er et al.}
\begin{document}

\title{X-ray, Optical and Infrared Observations of GX 339-4 During Its 2011 Decay}
\author{Tolga Din\c{c}er\altaffilmark{1}, Emrah Kalemci\altaffilmark{1}, Michelle M. Buxton\altaffilmark{2}, Charles D. Bailyn\altaffilmark{2}, John A. Tomsick\altaffilmark{3}, and Stephane Corbel\altaffilmark{4}}
\affil{}
\altaffiltext{1}{Sabanc{\i} University, Orhanl{\i}-Tuzla, 34956, \.{I}stanbul, Turkey}
\altaffiltext{2}{Astronomy Department, Yale University, P.O. Box 208101, New Haven, CT 06520-8101, USA}
\altaffiltext{3}{Space Sciences Laboratory, 7 Gauss Way, University of California, Berkeley, CA 94720-7450, USA}
\altaffiltext{4}{Universit\'{e} Paris 7 Denis Diderot and Service d'Astrophysique, UMR AIM, CEA Saclay, F-91191 Gif sur Yvette, France}

\begin{abstract}
We report multiwavelength observations of the black hole transient GX 339-4 during its outburst decay in 2011 using the data from \textit{RXTE, Swift} and SMARTS. Based on the X-ray spectral, temporal, and the optical/infrared (OIR) properties, the source evolved from the soft-intermediate to the hard state. Twelve days after the start of the transition towards the hard state, a rebrightening was observed simultaneously in the optical and the infrared bands. Spectral energy distributions (SED) were created from observations at the start, and close to the peak of the rebrightening. The excess OIR emission above the smooth exponential decay yields flat spectral slopes for these SEDs. Assuming that the excess is from a compact jet, we discuss the possible locations of the spectral break that mark the transition from optically thick to optically thin synchrotron components. Only during the rising part of the rebrightening, we detected fluctuations with the binary period of the system. We discuss a scenario that includes irradiation of the disk in the intermediate state, irradiation of the secondary star during OIR rise and jet emission dominating during the peak to explain the entire evolution of the OIR light curve.
\end{abstract}


\section{Introduction}
Galactic black hole transients are binary systems that undergo sporadic outbursts that last for months. During an outburst, they exhibit two main X-ray spectral states: the hard state and the soft state. In the soft state the X-ray energy spectra are dominated by a thermal disk component, and weak or no variability is observed; whereas in the hard state, the X-ray energy spectra are dominated by a non-thermal component, and high variability is observed. There also exists intermediate states in which the X-ray properties of the systems do not suit to the main states, but present the properties of a mixture of both states (see \citealt{Belloni10} for the details of the spectral states). Besides the correlated X-ray spectral and temporal properties, black hole transients also show state dependent radio, optical and infrared (OIR) properties. In the soft state, the radio emission is quenched \citep{Fender99, Russell_q11} indicating the jet turning off. In the hard state, compact steady jets are observed \citep{Fender06}. Therefore, from observational point of view the transitions from the soft to the hard state serve the perfect conditions to study the properties of the accretion flow for jet formation. Also in the hard state, the X-ray flux is positively correlated with both the radio flux \citep{Corbel00, Corbel03, Gallo03} and the OIR flux \citep{Russell06}. These relations suggest a common origin between the jet emission and X-rays. Some jet models assume that a hot electron corona is the base of the jet to explain these relations \citep{Markoff03}. There are also reports of direct jet synchrotron emission dominating the entire X-ray emission in the hard state at $10^{-4}<L_{\rm Edd}<10^{-3}$ \citep{Russell10}. Softening of the X-ray spectrum observed in some sources \citep{Tomsick01, Corbel06, Dincer08, Sobolewska11} at very low flux levels also suggests a change in the X-ray emission mechanism.

The X-ray observations of GX 339-4 in January 2010 revealed the start of an outburst \citep{Yamaoka2380, Tomsick2384}. Its multiwavelength observations during the rise and spectral properties during the state transitions have been reported elsewhere \citep{Cadolle11, Gandhi11, Shidatsu11_soft, Shidatsu11_hard, Stiele11, Yan11}. In January 2011, the source made a transition from the intermediate to the hard state during the outburst decay \citep{Munoz3117}. The optical observations through the end of February 2011 revealed a rebrightening \citep{Russell3191} which was also observed in previous outburst decays of GX 339-4 \citep{Buxton12}. Here, we report the results of \textit{RXTE, Swift}, and SMARTS observations of GX 339-4 in 2011 decay. We characterize the evolution of the X-ray spectral, temporal, and OIR photometric properties, and produce spectral energy distributions (SED) in order to investigate the jet-disk connection.

\section{Observations and Analysis}
\subsection{\textit{RXTE} Observations}
The outburst decay was amply covered with 54 pointed \textit{RXTE} observations between MJD 55,560 and 55,650 (2010 December 30 and 2011 March 30). However, some observations were not statistically satisfactory due to the short good time intervals (GTI) and/or small number of Proportional Counter Unit (PCU) on during the operation. Therefore, we did not include the observations with GTI $<$ 500 s (see Table~\ref{paramt} for a log of observations). 

We used data from the Proportional Counter Array (PCA) instrument onboard the \textit{RXTE} for the spectral analysis \citep{Jahoda96}. In most of the observations, the spectra were extracted in the 3-25 keV energy band, but in a few cases for which the noise dominated above 20 keV, we used the 3-20 keV band. The response matrix, and the background model were created using the standard FTOOLS (v6.11) programs. We added 0.5\% systematic error to the energy spectra following the suggestions of the \textit{RXTE} team.

The spectral analysis were performed using XSPEC 12.0.7 \citep{Arnaud96}. We employed a spectral model for the continuum that consists of absorption, a multicolour disk black-body and a power law. We also included a phenomenological smeared edge model \citep{Ebisawa94} for the iron $K_{\alpha}$ absorption edge seen around 7.1 keV to obtain acceptable $\chi^2$ values for the observations before MJD 55,606. In the spectral fits, the hydrogen column density, $N_{H}$, and the smeared edge width were fixed at $0.5\times10^{22}~\rm cm^{-2}$ \citep{Kong00} and 10 keV, respectively. This model was used previously in \cite{Tomsick01,Kalemci04,Kalemci05,Kalemci06}.

The Galactic ridge emission was an important factor for the faint observations. In order to estimate its spectrum we compared quasi-simultaneous \textit{RXTE} and \textit{Chandra} observations obtained on MJD 52,911.  We combined seven RXTE/PCA observations taken on the same day and fitted with a model consisting of interstellar absorption, a power law and a Gaussian to represent the Galactic ridge emission, and a second power law to represent the source. The centroid energy and width of Gaussian was fixed at 6.6 keV and 0.5 keV respectively. The parameters of the second power law from the source was set to the values obtained from \textit{Chandra} observation \citep{Gallo03}. With this method we modelled the Galactic ridge emission with a power law index of 2.1 and an unabsorbed flux of 7.55 $\times$ 10$^{-12}$ ergs cm$^{-2}$ s$^{-1}$ in the 3-25 keV band \citep{Dincer08, Coriat09}. We applied Galactic ridge emission correction to the spectra, fluxes and rms amplitudes of the observations after MJD 55,627. The contamination from Galactic ridge emission was not greater than 3\% of the total flux for the observations before this date. 

For each PCA observation, we produced power density spectrum (PDS) from segments of length 256 s with a time resolution of 2$^{-9}$ s in 3-30 keV energy band using IDL programs developed at University of T\"{u}bingen. The averaged PDS were normalized as described in \cite{Miyamoto89} and corrected for the dead time effects according to \cite{Zhang95}. Then, PDS were fit using Lorentzians in 0.004-256 Hz range. The rms amplitudes are obtained by integrating the normalized PDS and corrected for both the background and the ridge emission as described in \cite{Kalemci06}. All spectral and timing results are presented in Table~\ref{paramt}.

\subsection{\textit{Swift} Observations}
We also analyzed \textit{Swift} X-ray Telescope (XRT) observations conducted at the same time period with the \textit{RXTE} observations. We found 12 observations carried out between MJD 55,622 and 55,647. We used them together with the \textit{RXTE} observations and looked for presence of any spectral softening.

We analyzed the XRT photon counting mode event data using XRTPIPELINE task provided in FTOOLS package. Pile up was an issue for the first four observations whose count rate were greater than 1 c/s. To remove its effects, following the SWIFT SCIENCE DATA CENTER (SSDC) recommendations, we selected the source photons from a ring with an inner radius of 5$\arcsec$ and an outer radius of 40$\arcsec$. For the rest of the observations, the source photons were selected in a circular region with a radius of 40$\arcsec$. The background photons were accumulated from a ring with an inner radius of 70$\arcsec$ and an outer radius of 100$\arcsec$ centered at the source position. 

For the spectral analysis, only events with grades 0-12 were selected. The auxiliary response files were created by XRTMKARF and corrected using the exposure maps, and the standard response matrix swxpc0to12s6\_20010101v013.rmf was used. We binned the energy spectra by fixing the number of counts per bin at 50. We fitted the spectra with a model that consists of photoabsorption and a power law in 0.6-8.0 keV band. In our initial spectral runs, we let the $N_{H}$ free, and the resulting values were between $(0.3-0.7)\times10^{21}~\rm cm^{-2}$. As we were not able to constrain the $N_{H}$, we performed a second run with $N_{H}$ fixed at $0.5\times10^{22}~\rm cm^{-2}$. The log of \textit{Swift} observations and the spectral results are presented in Table~\ref{paramt}.

\subsection{SMARTS Observations}
The regular optical/infrared observations were performed with the ANDICAM \citep{Depoy03} camera on the SMARTS 1.3m telescope in $V$, $I$, $J$ and $H$ bands. The observations covered the outburst decay in daily basis between MJD 55,582 and 55,720. In this paper, we focus on the OIR light curves and evolution of the spectral energy distributions (SED). The dereddening of the observed magnitudes and their conversion to physical units were critical to create SEDs. For this purpose, we used the optical extinction, $A_V=3.7 \pm 0.3$ \citep{Zdziarski98} together with the extinction laws given by \cite{ccm89} and \cite{Odonnell94}. The same $A_V$ was previously utilized in \citealt{Corbel02, Coriat09, Buxton12} for the SED creation. For the details of the selection of $A_V$, dereddening and the flux conversion procedures, we refer to \cite{Buxton12}.

\begin{deluxetable*}{ccccccccc}
\tablewidth{0pt}
\tablecaption{Observational Parameters Obtained From RXTE Data}
\tablehead{
\colhead{Observation} & \colhead{} & \colhead{GTIs} &
\colhead{} & \colhead{$T_{in}$} & \colhead{} & \colhead{} & \colhead{rms\tablenotemark{e}} & \colhead{$\nu$\tablenotemark{f}}\\ 
\colhead{ID.\tablenotemark{a}} & \colhead{MJD\tablenotemark{b}}     & \colhead{(ksec)} &
\colhead{$\Gamma$} & \colhead{(keV)} &
\colhead{PL Flux\tablenotemark{c}} & \colhead{DBB Flux\tablenotemark{d}} &
\colhead{(\%)} & \colhead{(Hz)}}
\startdata
45-00 &  55,559.58 & 0.77 & 2.44 $\pm$ 0.11 & 0.60 $\pm$ 0.01 &  4.45 &  8.02 & $<$1.82 & \nodata\\
01-00 &  55,561.05 & 0.70 & 2.46 $\pm$ 0.09 & 0.58 $\pm$ 0.01 &  5.49 &  7.35 & $<$2.75 & \nodata\\
01-01 &  55,563.14 & 0.74 & 2.34 $\pm$ 0.14 & 0.59 $\pm$ 0.02 &  4.14 &  7.33 & $<$5.90 & \nodata\\
01-02 &  55,565.82 & 0.53 & 2.46 $\pm$ 0.13 & 0.57 $\pm$ 0.02 &  5.63 &  5.98 & $<$6.08 & \nodata\\
01-03 &  55,567.91 & 0.60 & 2.60 $\pm$ 0.07 & 0.58 $\pm$ 0.02 &  8.71 &  5.30 & $<$2.73 & \nodata\\
02-03 &  55,574.87 & 1.03 & 2.27 $\pm$ 0.09 & 0.56 $\pm$ 0.01 &  5.95 &  4.37 & $<$5.25 & \nodata\\       
03-00 &  55,576.85 & 1.44 & 2.50 $\pm$ 0.07 & 0.56 $\pm$ 0.01 &  7.51 &  4.02 & $<$4.15 & \nodata\\       
03-01 &  55,578.88 & 1.40 & 2.40 $\pm$ 0.06 & 0.56 $\pm$ 0.01 &  6.94 &  3.59 & 10.16 $\pm$ 1.71 & \nodata\\
03-02 &  55,580.61 & 1.32 & 2.45 $\pm$ 0.06 & 0.55 $\pm$ 0.02 &  8.05 &  3.62 & 8.08 $\pm$ 0.53 & \nodata\\                 
04-00 &  55,582.70 & 0.54 & 2.57 $\pm$ 0.12 & 0.54 $\pm$ 0.03 &  7.32 &  3.09 & $<$7.50 & \nodata\\			
04-04 &  55,585.94 & 0.98 & 2.32 $\pm$ 0.04 & 0.56 $\pm$ 0.01 &  7.62 &  3.20 & 8.88 $\pm$ 0.19 & 2.09 $\pm$ 0.06\\			
04-02 &  55,586.49 & 0.70 & 2.46 $\pm$ 0.10 & 0.55 $\pm$ 0.03 &  7.56 &  2.96 & 8.52 $\pm$ 0.73 & \nodata\\
04-07 &  55,587.50 & 0.52 & 2.38 $\pm$ 0.12 & 0.54 $\pm$ 0.03 &  6.38 &  2.35 & 7.82 $\pm$ 1.00 & \nodata\\
04-08 &  55,588.55 & 0.54 & 2.36 $\pm$ 0.14 & 0.53 $\pm$ 0.03 &  4.95 &  2.53 & $<$10.94 & \nodata\\
05-00 &  55,589.20 & 0.58 & 2.35 $\pm$ 0.12 & 0.52 $\pm$ 0.03 &  5.97 &  2.30 & $<$9.90 & \nodata\\
05-04 &  55,590.43 & 0.58 & 2.43 $\pm$ 0.12 & 0.51 $\pm$ 0.04 &  6.10 &  1.87 & $<$10.61 & \nodata\\
05-01 &  55,591.61 & 1.31 & 2.15 $\pm$ 0.05 & 0.53 $\pm$ 0.02 &  6.69 &  2.59 & 8.63 $\pm$ 0.25  & 1.77 $\pm$ 0.06 \\
05-05 &  55,592.73 & 0.83 & 2.35 $\pm$ 0.10 & 0.51 $\pm$ 0.04 &  6.25 &  1.67 & 11.30 $\pm$ 0.87 & 1.72 $\pm$ 0.05 \\
05-02 &  55,593.50 & 1.55 & 2.44 $\pm$ 0.08 & 0.52 $\pm$ 0.02 &  5.89 &  2.01 & 9.83 $\pm$ 0.43 & \nodata\\
05-03 &  55,594.89 & 1.23 & 2.29 $\pm$ 0.06 & 0.50 $\pm$ 0.05 &  7.31 &  1.10 & 17.54 $\pm$ 2.56 & \nodata\\
06-00 &  55,597.25 & 0.98 & 2.01 $\pm$ 0.03 & 0.42 $\pm$ 0.05 &  9.27 &  0.48 & 19.38 $\pm$ 1.29 & 1.03 $\pm$ 0.05 \\
06-01 &  55,598.66 & 1.09 & 2.11 $\pm$ 0.06 & 0.44 $\pm$ 0.08 &  7.94 &  0.50 & 25.18 $\pm$ 1.48 & \nodata\\
06-02 &  55,601.88 & 1.80 & 1.89 $\pm$ 0.03 & 0.44 $\pm$ 0.18 &  8.35 &  0.11 & 22.94 $\pm$ 1.71 & 1.39 $\pm$ 0.05\\
07-00 &  55,603.98 & 0.74 & 1.77 $\pm$ 0.04 & \nodata &  7.97 &  0.00 & 17.46 $\pm$ 1.62 & \nodata\\
07-03 &  55,604.89 & 1.28 & 1.75 $\pm$ 0.02 & \nodata &  7.61 &  0.00 & 26.16 $\pm$ 1.93 & \nodata\\
07-01 &  55,606.89 & 1.05 & 1.67 $\pm$ 0.04 & \nodata &  6.82 &  0.00 & 25.02 $\pm$ 1.12 & \nodata\\
07-02 &  55,607.76 & 1.26 & 1.71 $\pm$ 0.02 & \nodata &  6.05 &  0.00 & 23.04 $\pm$ 1.96 & \nodata\\
07-04 &  55,609.84 & 1.24 & 1.66 $\pm$ 0.02 & \nodata &  4.82 &  0.00 & 33.98 $\pm$ 2.58 & \nodata\\
08-00 &  55,611.60 & 1.44 & 1.70 $\pm$ 0.02 & \nodata &  3.74 &  0.00 & 37.76 $\pm$ 4.00 & \nodata\\
08-02 &  55,613.72 & 0.66 & 1.70 $\pm$ 0.03 & \nodata &  2.88 &  0.00 & 29.08 $\pm$ 1.93 & \nodata\\
08-01 &  55,615.45 & 0.85 & 1.64 $\pm$ 0.03 & \nodata &  2.50 &  0.00 & 32.84 $\pm$ 1.68 & \nodata\\
08-03 &  55,616.57 & 0.58 & 1.70 $\pm$ 0.04 & \nodata &  2.33 &  0.00 & 37.62 $\pm$ 8.70 & \nodata\\
09-00 &  55,617.53 & 1.68 & 1.69 $\pm$ 0.04 & \nodata &  2.16 &  0.00 & 40.49 $\pm$ 8.33 & \nodata\\
10-02 &  55,628.65 & 1.54 & 1.77 $\pm$ 0.07 & \nodata &  1.03 &  0.00 & 33.64 $\pm$ 2.97 & \nodata\\
10-03 &  55,630.23 & 0.70 & 1.70 $\pm$ 0.04 & \nodata &  1.05 &  0.00 & 35.26 $\pm$ 6.05 & \nodata\\
11-00 &  55,632.05 & 1.22 & 1.77 $\pm$ 0.09 & \nodata &  0.91 &  0.00 & 44.79 $\pm$ 4.17 & \nodata\\
11-02 &  55,636.20 & 0.59 & 1.68 $\pm$ 0.10 & \nodata &  0.70 &  0.00 & 43.23 $\pm$ 5.35 & \nodata\\
12-00 &  55,638.73 & 1.46 & 1.69 $\pm$ 0.12 & \nodata &  0.59 &  0.00 & 41.03 $\pm$ 3.58 & \nodata\\
12-01 &  55,639.50 & 1.52 & 1.86 $\pm$ 0.14 & \nodata &  0.50 &  0.00 & 51.31 $\pm$ 10.72 & \nodata\\
13-00 &  55,646.29 & 1.23 & 1.38 $\pm$ 0.24 & \nodata &  0.31 &  0.00 & \nodata & \nodata\\
13-01 &  55,649.64 & 0.62 & 1.72 $\pm$ 0.35 & \nodata &  0.28 &  0.00 & \nodata & \nodata\\
\cutinhead{Observational Parameters Obtained From Swift Data}
00031931011 & 55,622.66 & 1.20 & 1.54 $\pm$ 0.10 \\
00030943021 & 55,624.06 & 1.27 & 1.55 $\pm$ 0.10 \\
00030943022 & 55,626.00 & 1.20 & 1.47 $\pm$ 0.12 \\
00030943023 & 55,628.81 & 1.31 & 1.70 $\pm$ 0.12 \\
00031931012 & 55,629.82 & 1.16 & 1.54 $\pm$ 0.17 \\
00030943024 & 55,630.69 & 1.30 & 1.52 $\pm$ 0.10 \\
00030943025 & 55,632.43 & 1.06 & 1.65 $\pm$ 0.21 \\
00030943026 & 55,634.10 & 1.19 & 1.59 $\pm$ 0.16 \\
00030943027 & 55,638.93 & 1.30 & 1.58 $\pm$ 0.14 \\
00030943029 & 55,642.46 & 1.06 & 1.76 $\pm$ 0.19 \\
00031931014 & 55,643.21 & 1.19 & 1.65 $\pm$ 0.16 \\
00030943030 & 55,646.15 & 2.20 & 1.59 $\pm$ 0.13 \\

\enddata
\tablenotetext{a}{Full observation ID is, 95409-01-Obs for the first observation, and 96409-01-Obs for the rest.}
\tablenotetext{b}{Modified Julian Date (JD$-$2,400,000.5) at the start of the observation.}
\tablenotetext{c}{Unabsorbed power law flux in the 3-25 keV band, in units of $10^{-10}$ ergs $\rm cm^{-2}$ $\rm  s^{-1}$.}
\tablenotetext{d}{Unabsorbed disk black-body in the 3-25 keV band, in units of $10^{-10}$ ergs $\rm cm^{-2}$ $\rm  s^{-1}$.}
\tablenotetext{e}{Total rms amplitude of variability integrated over a range of 0-$\infty$ Hz in the 3-30 keV band.}
\tablenotetext{f}{QPO centroid frequency}
\label{paramt}
\end{deluxetable*}

\section{Results}

\begin{figure}
\epsscale{1.17}
\plotone{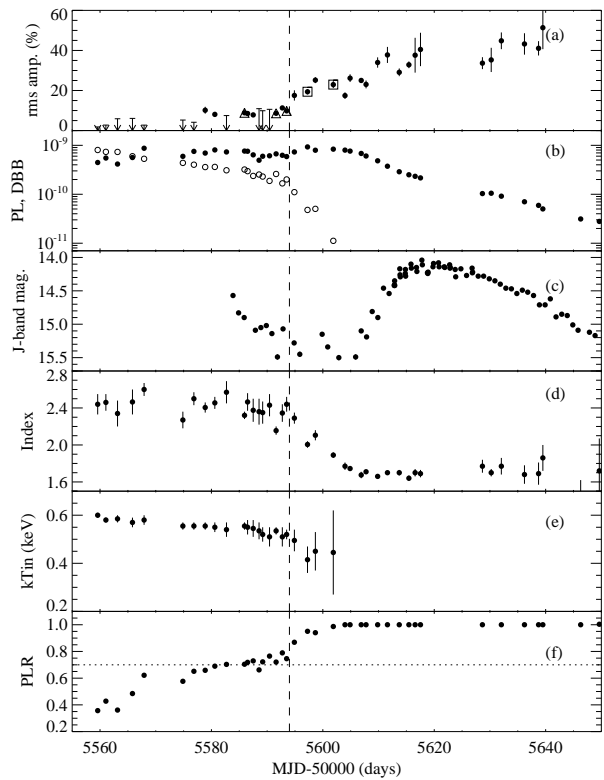}
\caption{Evolution of (a) the total rms amplitude of variability in the 3-30 keV band, (b) the power law flux (filled circles) and the disk-blackbody flux (empty circles) in the 3-25 keV band in units of $10^{-10}$ ergs $\rm cm^{-2}$ $\rm s^{-1}$, (c) J-band IR light curve, (d) the photon index $\Gamma$, (e) the inner disk temperature $T_{in}$, (f) the ratio of the power law flux to the total flux in the 3-25 keV band. Dashed line indicates the time of state transition on MJD 55,594. Triangles and squares show the observations with type B and C QPOs, respectively.}
\label{params}
\end{figure}

\subsection{X-ray Evolution}
\label{SecXevol}
\begin{figure}
\epsscale{1.17}
\plotone{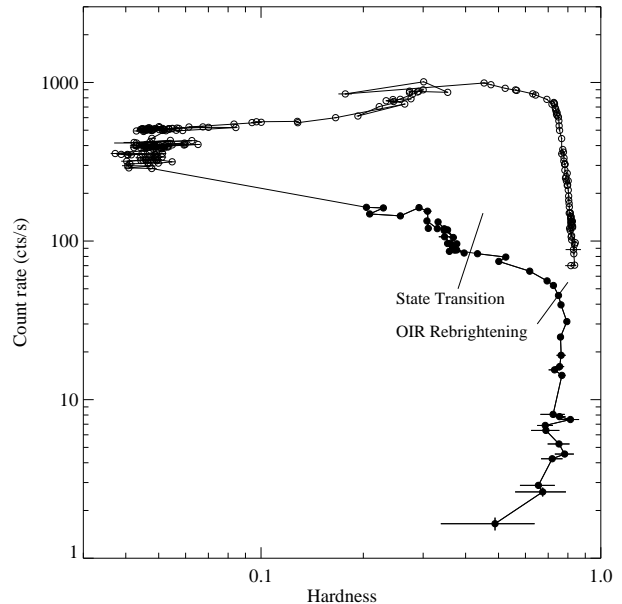}
\caption{Start of the state transition and the approximate time of the OIR rebrightening are marked in the hardness-intensity diagram of the entire outburst.  Observations analyzed in this paper are shown with filled circles (colored red in the online version).}
\label{HI}
\end{figure}

In Figure~\ref{params}, we present the evolution of the spectral and the temporal parameters. On MJD 55,594, dramatic changes occurred in both the X-ray spectral and the temporal parameters. The rms amplitude of variability jumped from 9.8\% to 17.4\% in one day. Both the disk flux and the temperature of the inner disk decreased. The power law flux increased and the photon index started to harden. These changes in the evolution of the parameters suggest a reshaping of the accretion dynamics and lead to a transition from the soft-intermediate state towards the hard state. In Figure~\ref{HI}, we show the hardness-intensity diagram and mark the start of the state transition. Based on the extrapolation of the power law model, the X-ray luminosity at the start of the transition is $L_{\rm 1-200~keV}$ = 2.7 $\times$ $10^{37}$ ergs s$^{-1}$ or $\sim$ 2\% $L_{\rm Edd}$ if we adopt a distance of 8 kpc \citep{Hynes04, Zdziarski04} and a mass of 10 $\Msun$ \citep{Hynes03, Munoz08}.

Before the transition, the energy spectra were soft with a mean photon index of 2.37. There was comparable contribution to the total X-ray flux from the disk and the power law components. The power law ratio (PLR, the ratio of the power law flux to the total flux in the 3-25 keV band) was increasing from 0.40 to 0.75 level. This increase was due to the steady decrease in the disk flux. In addition to the decreasing disk flux, the inner disk temperature was also decreasing. The first detection of the X-ray variability occurred when the power law ratio reached 0.7. The rms amplitude of variability was less than 10\%. Our analysis confirms the type B QPO detections reported in \citealt{Stiele11}. All these spectral properties indicate that the source was in transition from the soft to the hard state, or simply in the intermediate state \citep{Kalemci04}.

After the transition, the energy spectra became dominated by the power law component in six days. At the same time, the photon index hardened from 2.3 to 1.8. The power law flux increased, and remained at a higher level than its soft-intermediate state level. Again during this six days the rms amplitude of variability increased to 25\%. For two observations, Type C QPOs (according to the classification in \cite{Motta11}) were detected (see Figure~\ref{params} and Table~\ref{paramt}). After MJD 55,605 the disk component was no more significant and no longer needed in the energy spectra. The power law flux started to decay and the photon index kept decreasing until it leveled off at 1.70. The rms amplitude of variability gradually increased to 50\% in thirty days.

\begin{figure}
\epsscale{1.17}
\plotone{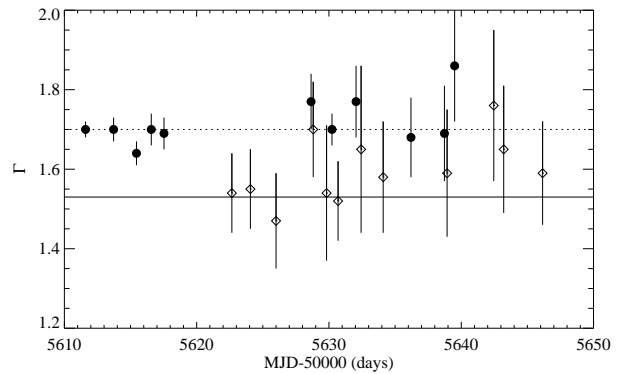}
\caption{Evolution of the photon index obtained from RXTE (circles) and Swift (diamonds) observations between
MJD 55,611 and 55,650. Best fit values at 1.53 and 1.70 constant levels for RXTE (dotted line) and Swift (solid line) only points, respectively.}
\label{softening}
\end{figure}

\subsection{No Evidence For Softening}
We also inspected the possible presence of softening of the spectra at low flux. In Figure~\ref{softening}, we plotted the evolution of the photon index obtained from both the RXTE and Swift observations between MJD 55,611 and 55,650. The RXTE indices are systematically higher than the Swift indices, however both data sets are separately consistent with a flat evolution ($\Gamma=1.53 \pm 0.06$, $1.70 \pm 0.01$ for Swift and RXTE, respectively). The reason for the systematic difference may be caused by the use of different energy bands in RXTE and Swift spectra. If the Galactic ridge emission is underestimated, the RXTE spectral indices would become slightly harder, but not enough to account for the entire difference. Regardless of the deviation between two data sets we conclude that the data suggests no evidence for the softening of the spectra between MJD 55,610 and 55,650.

\begin{figure}
\epsscale{1.16}
\plotone{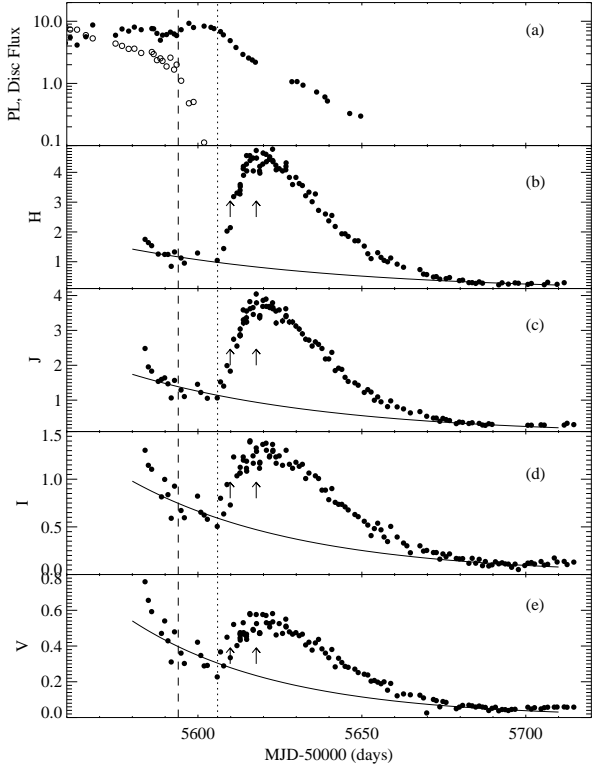}
\caption{RXTE/PCA X-ray, and SMARTS OIR light curves of GX 339-4. (a) X-ray flux in the 3-25 keV band in units of $10^{-10}$ ergs $\rm cm^{-2}$ $\rm s^{-1}$. Filled circles: power law flux (PL), empty circles: disk flux.  (b-e) Undereddened H, J, I and V light curves in units of mJy. (Error bars are smaller than the plot symbols.) The dashed line indicates the X-ray state transition and the dotted line indicates the start time of the OIR rebrightening. The solid lines show the baseline emission possibly originating in the disk. The arrows point the dates for which we constructed the SEDs  (MJD 55,609.84 and MJD 55,617.53).}
\label{mwlc}
\end{figure}

\subsection{Light Curves}
\label{Seclc}
In Figure~\ref{mwlc}, we present the evolution of the power law and disk black body fluxes together with the OIR light curves obtained during the 2011 decay. The dashed line shows the time of state transition from soft-intermediate to hard-intermediate state whereas the dotted line shows the start time of the OIR rebrightening. To find the start of the rebrightening, we first formed a baseline that smoothly connects the fluxes before (between MJD 55,590 - 55,604) and after (MJD 55,680 - 55,690) the flare as an exponential decay. We assumed that the physical origin of the rebrightening is separate from that of the baseline. We then fitted the rise of the rebrightening (between MJD 55,608 - 55,612) in the infrared bands only with a straight line over the baseline. We used the infrared, since the fluctuations are lower during the early part of the rebrightening compared to the optical bands. The start of the flare is defined as the date that the linear fit intersects zero, and it is MJD 55,607 $\pm$ 1 day.

The evolution of the disk and the power law flux were described in $\S$~\ref{SecXevol}. As the source transitioned to the hard-intermediate state the power law flux increased, and the OIR rebrightening occurred when the spectrum was almost its hardest (see also Figures~\ref{params} and \ref{HI}). There was a delay of $\sim$12 days between the increase in the power law X-ray flux and the OIR rebrightening. The rebrightening started at a PCA flux of $F_{\rm 3-25~keV}$=7.6 $\times~10^{-10}$ ergs cm$^{-2}$ s$^{-1}$, resulting in a bolometric luminosity of $L_{\rm 1-200~keV}$=1.8 $\times~10^{37}$ ergs s$^{-1}$ or $\sim$ 1.4\% $L_{\rm Edd}$. Note that such a delay has already been noticed for other black hole transients, namely XTE J1550-564 \citep{Kalemci06smqw}, 4U 1543-47 \citep{Kalemci05} and also for GX 339-4 \citep{Coriat09}.

The evolution in OIR in different bands are similar (see Figure~\ref{mwlc}). A decay is followed by a $\sim$70 days of rebrightening that peaked around the same dates before reaching a constant level. The amount of brightening is, however, different among the bands. Ratio of the peak flux to the baseline flux decreases from H to V (max $\sim$4.9 to $\sim$3 on MJD 55,620). 

\subsection{Evidence For Binary Period in The Optical Light Curves}
\label{Scperiod}
OIR light curves shown in Figure~\ref{mwlc} fluctuate during the initial decay (between MJD 55,580 - 55,605) in a time scale of days. The fluctuations continue even on the rise and the peak of the rebrightening. Moreover, some parts of the light curves seem to show regular modulations. Therefore, we decided to search for periodicity in all bands using the Lomb-Scargle algorithm \citep{Scargle82}. The initial decay (between MJD 55,580 - 55,605) did not provide a significant peak in the periodogram in any of the bands. Likewise, there is no evidence for periodicity in the light curves after MJD 55,630.

\begin{figure*}
\epsscale{1.16}
\plottwo{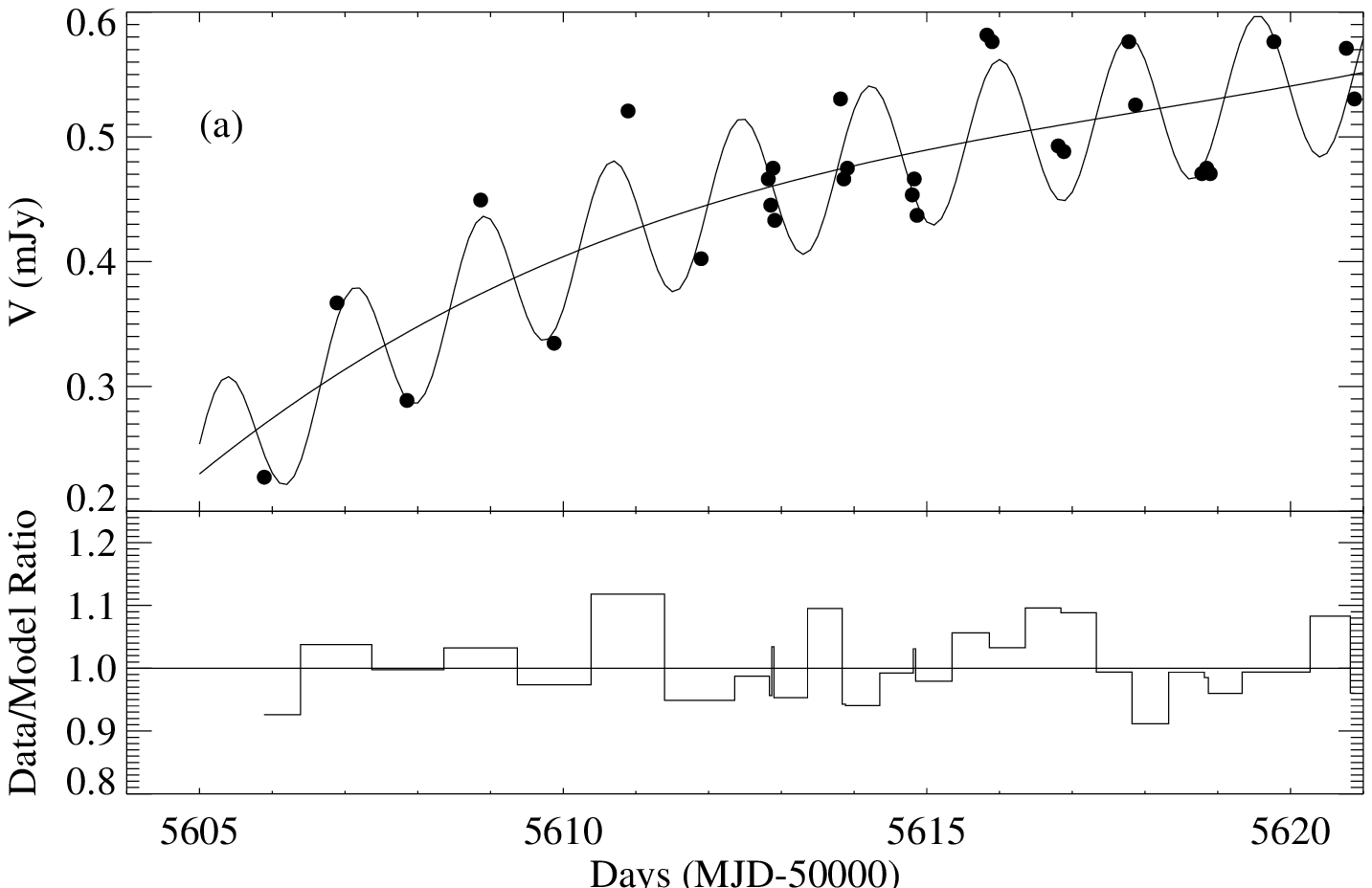}{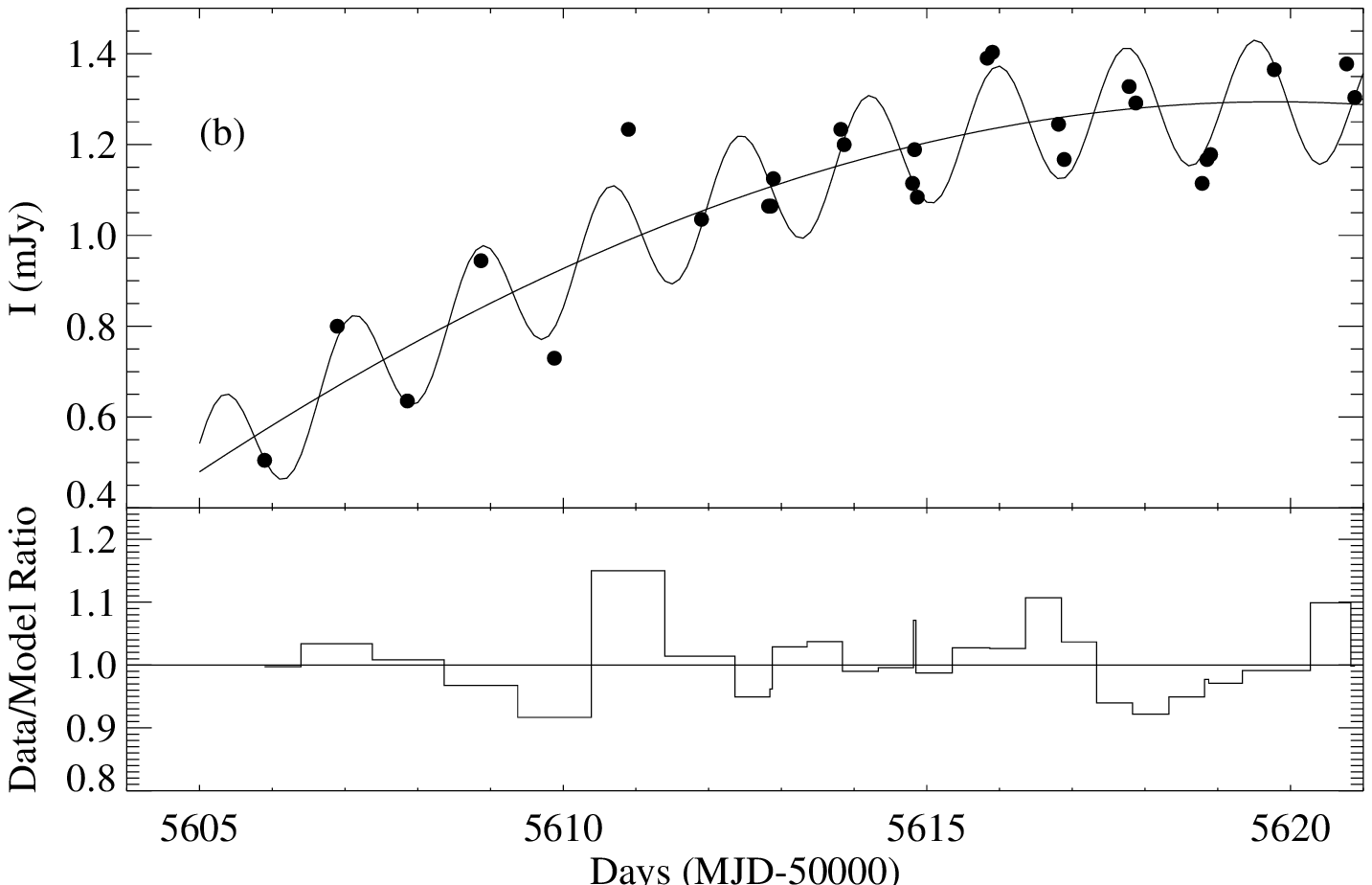}
\plottwo{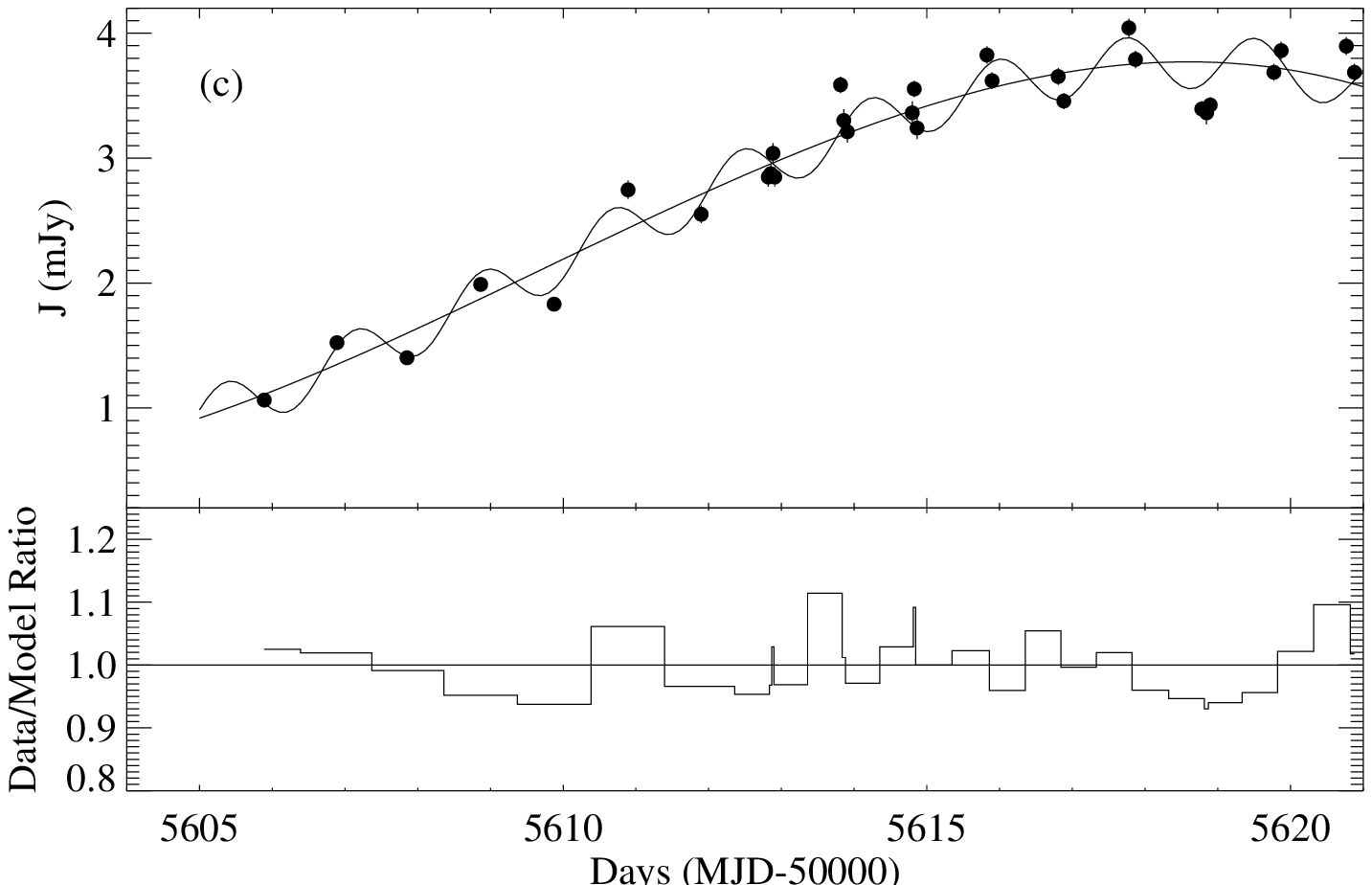}{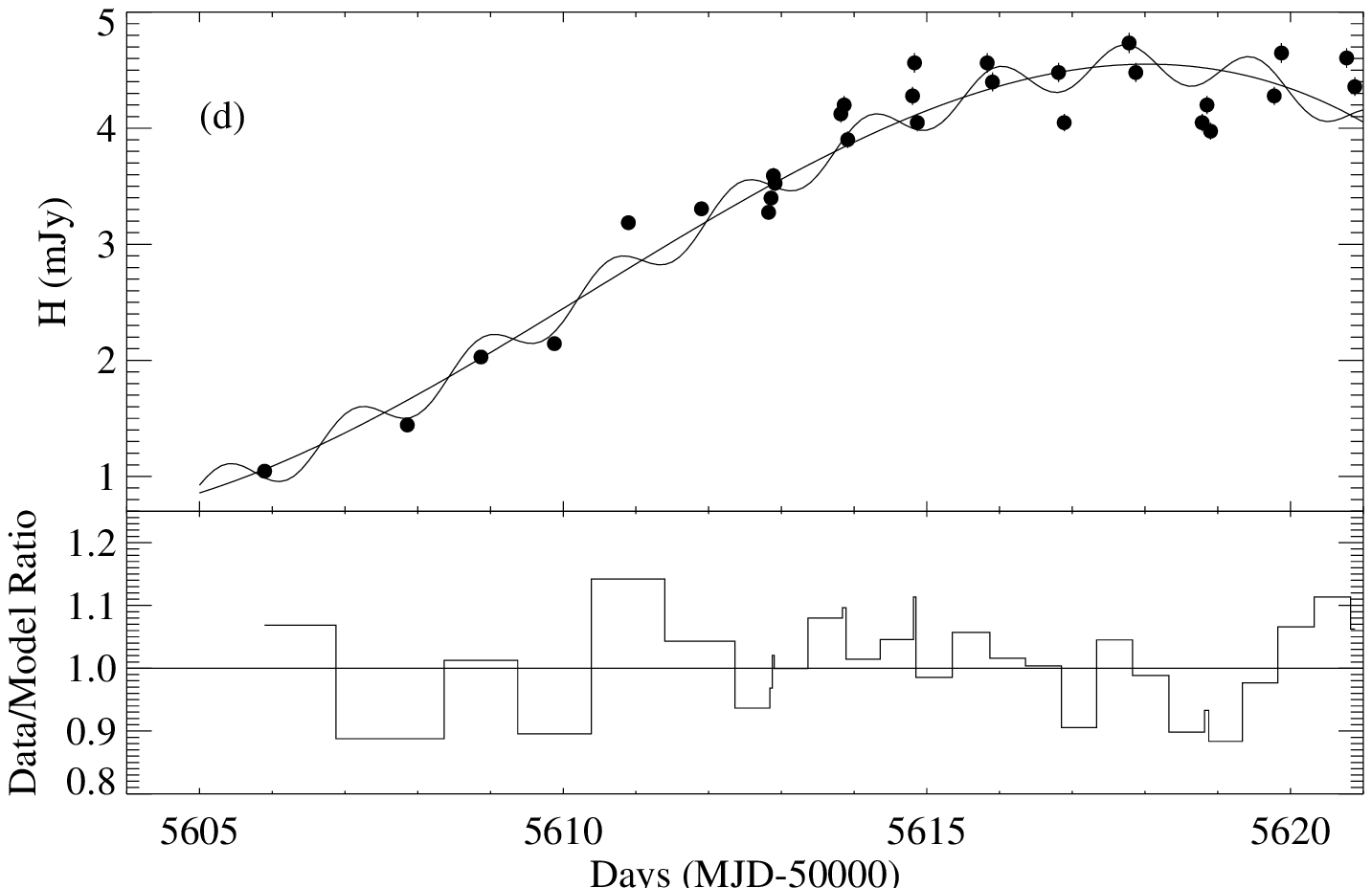}
\caption{Fluctuations in (a-d) V, I, J and H light curves respectively in flux units of mJy. The solid lines describe the rise of the rebrightening together with the fluctuations with a third degree of polynomial with a sinusoidal.}
\label{lcfluc} 
\end{figure*}

On the other hand, we detected the known binary period of the system at 1.77 days \citep{Hynes03,Levine06} in the V, I and J band light curves between MJD 55,605 and 55,621, which is the rise and the peak of the rebrightening. Using the method of \cite{Horne86} the false alarm probabilities of the known period are estimated as 2.2$ \times 10^{-4}$, 1.6$ \times 10^{-4}$, 7.6$ \times 10^{-3}$ (3.69-$\sigma$, 3.78-$\sigma$, 2.78-$\sigma$) for the V, I and J bands, respectively. We performed a detailed fitting with the following function:
\begin{equation}
F_{oir}=a+bt_0+ct_0^2+dt_0^3+A~sin(\frac{2\pi}{P}t_0+\phi)
\label{Eq}
\end{equation}
where $t_0$ is defined as the time from MJD 55,605, and $F_{oir}$ is in units of mJy. The polynomial part of the Eq.~\ref{Eq} represents the continuum of the rebrightening whereas the periodic part represents the modulations with the binary period P = 1.77 days obtained from the periodogram. The model parameters obtained from the fits are given in Table~\ref{lcfitpar} and the best model fits to the data are shown in Figure~\ref{lcfluc}. The phases are consistent among all bands. Note that for the H band the periodic modulation is not necessary to fit the data, however the first few points are consistent with the binary period if the modulation is included in the fit.

\begin{deluxetable*}{lcccc}
\tablecaption{Model Parameters Derived From The Light Curves (for a fixed period $P=1.77$ days)}
\tablehead{\colhead{Parameters} & \colhead{V} & \colhead{I} & \colhead{J} & \colhead{H}}
\startdata
$a~(mJy)$ & 0.229 $\pm$ 0.001 & 0.479 $\pm$ 0.007 & 0.918 $\pm$ 0.017 & 0.820 $\pm$ 0.050 \\
$b~(mJy~d^{-1})$ & 0.047 $\pm$ 0.002 & 0.105 $\pm$ 0.004 & 0.200 $\pm$ 0.029 & 0.213 $\pm$ 0.026 \\
$c~(mJy~d^{-2})$ & -0.0029 $\pm$ 0.0003 & -0.0031 $\pm$ 0.0006 & 0.017 $\pm$ 0.004 & 0.033 $\pm$ 0.004 \\
$d~(mJy)$ & (7.23 $\pm$ 1.25) $\times$ 10$^{-5}$ & (-2.27 $\pm$ 2.62) $\times$ 10$^{-5}$ & -0.0011 $\pm$ 0.0002 & -0.0021 $\pm$ 0.0001 \\
$A~(mJy)$ & 0.061 $\pm$ 0.001 & 0.136 $\pm$ 0.003 & 0.216 $\pm$ 0.02 & 0.17 $\pm$ 0.02 \\
$\phi~(rad)$ & 0.40 $\pm$ 0.02 & 0.48 $\pm$ 0.02 & 0.320 $\pm$ 0.079 & 0.232 $\pm$ 0.105 \\
\enddata
\label{lcfitpar}
\end{deluxetable*}

\begin{figure}
\epsscale{1.17}
\plotone{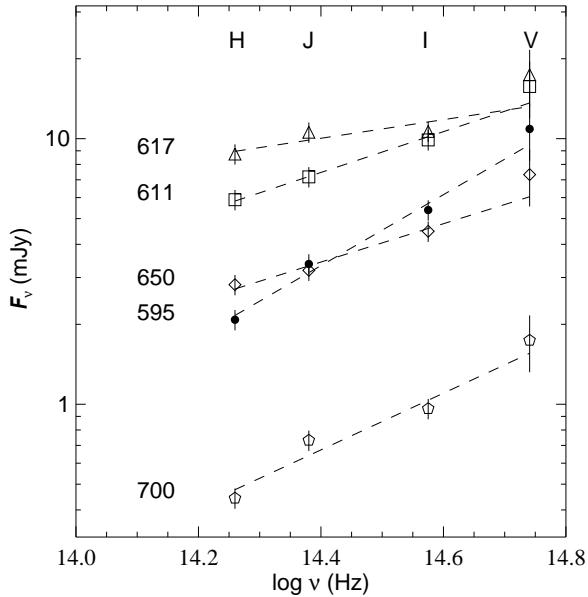}
\caption{Dereddened OIR SEDs sampled from different stages of both intermediate and hard states. The numbers indicate the dates in the form of MJD-55,000. The dashed lines show the power law best fits to the data from individual days.}
\label{SEDall}
\end{figure}

\begin{figure*}
\epsscale{1.17}
\plottwo{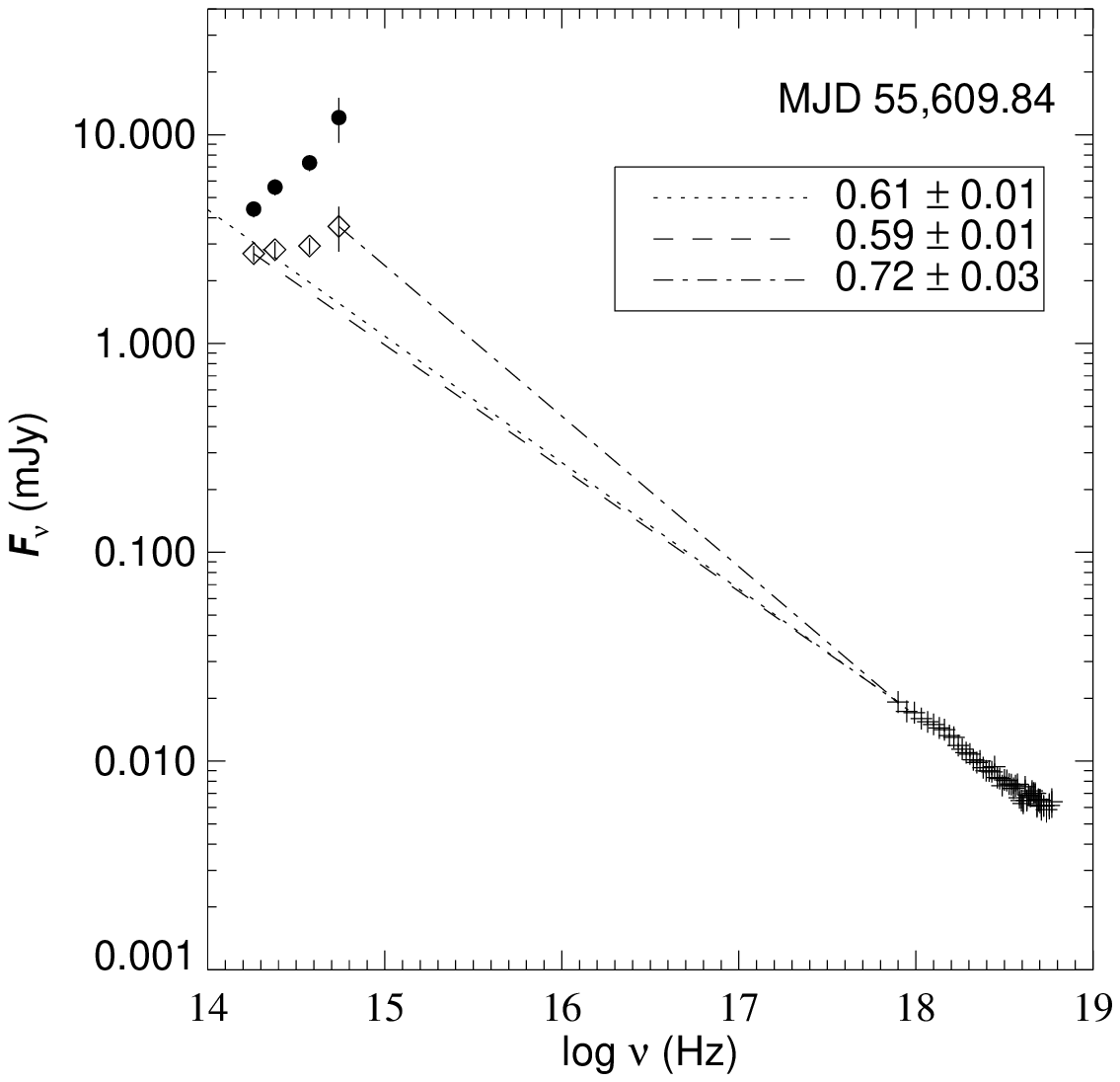}{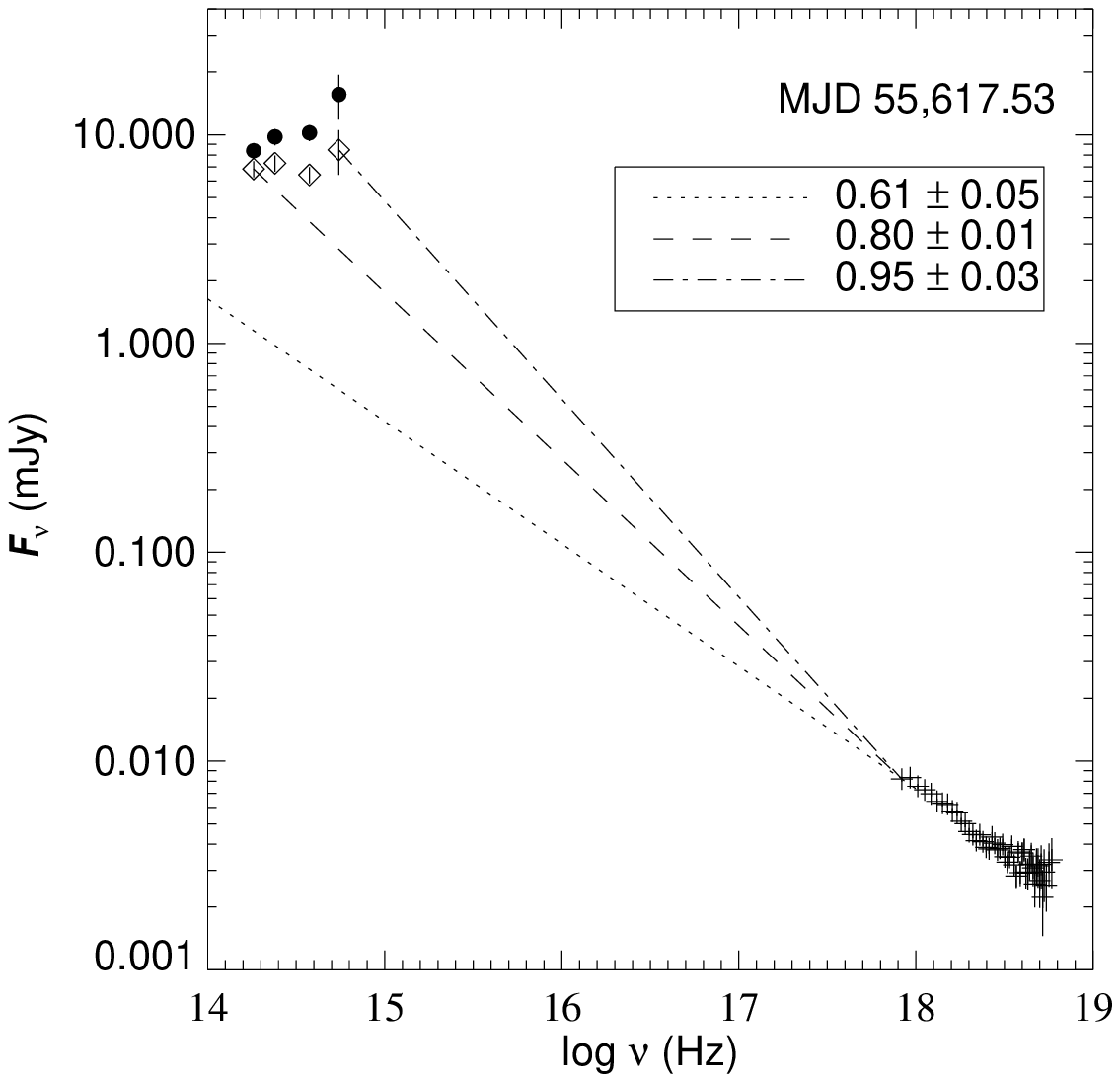}
\caption{Broadband SEDs for MJD 55,609.84 and MJD 55,617.53. Black circles: total flux density of the source, diamonds: baseline subtracted flux density. Dotted line: power law fit extrapolations from X-rays, dashed line: power law fit from H-band to softest X-rays (3 keV), dashed-dotted lines: power law fit from V-band to softest X-rays (3 keV). The power law indices from the fits are shown in the legends. The errors on the slopes are calculated from the difference between the two slopes obtained by the fits from mid points of the OIR flux densities and the 1~$\sigma$ above the mid points.}
\label{seds}
\end{figure*}

\subsection{SEDs}
We created spectral energy distributions to inspect the physical components and the mechanisms that produce the observed emission in both the intermediate state and the hard state. Below we will present our observational results by exploring the evolution of the OIR SED alone, and then together with the X-rays.

\subsubsection{OIR SEDs}
Figure~\ref{SEDall} shows several SEDs of the total emission sampled from different stages of the OIR evolution. In order to quantitatively track the evolution, each SED was fitted with a power law. The evolution of the SED started with higher spectral slopes (all slopes are negative, and $\alpha$ represents the absolute value of the slopes throughout the paper) in the hard intermediate state ($\alpha=1.33$ on MJD 55,595) and continuously decreased ($\alpha=0.76$ on MJD 55,611) until the peak of the OIR rebrightening in the hard state. During the peak, the slope of the SED became flatter ($\alpha=0.35$ on MJD 55,618). As the OIR decayed, the slope started to increase again ($\alpha=0.72$ on MJD 55,650). At the end of the decay where the OIR emission decreased to a constant level in all bands, the spectral slope had a less steep value ($\alpha=1.07$ on MJD 55,700) than it was in the hard intermediate state. The typical errors for the spectral slopes were around $\pm$ 0.15 at 1-$\sigma$, but note that  the relative slopes are not affected by the dereddening process which dominates the error calculation.

\subsubsection{Broadband SEDs}
\label{SCbbsed}
In order to understand the emission mechanism that gives rise to the rebrightening, we constructed two broadband SEDs in Figure~\ref{seds} with the total OIR (circles), the excess OIR (diamonds), and the X-ray fluxes from MJD 55,609.84 and 55,617.53 (rise and peak of rebrightening, see Figure~\ref{mwlc}). The total OIR fluxes were calculated using the polynomial part of Equation 1 (the solid curves in Figure~\ref{lcfluc}) This way we removed the effect of modulations with the binary period found in $\S$~\ref{Scperiod}. We then removed the baseline flux to find the SED of the excess. We calculated the errors on the flux densities by only considering the calibration and interstellar extinction. The errors from the baseline and the rebrightening fits were negligable. The excess OIR spectra yielded flat spectra with spectral slopes of 0.15 $\pm$ 0.15 for MJD 55,609.84 and -0.05 $\pm$ 0.15 for MJD 55,617.53.  We refer to the Discussion section for the explanation of the lines in Figure~\ref{seds}.

\section{Discussion}
The changes observed in both the spectral and the temporal parameters imply a change in accretion dynamics. The most clear changes in both the X-ray spectral and temporal parameters occurred on MJD 55,594. We marked this date as the start of the transition from soft-intermediate state towards the hard state. Throughout the the decay phase of the outburst, GX 339-4 generally presented typical X-ray and OIR behaviour as observed from other black holes, and also of its previous outburst decays.

\subsection{Evolution In The X-ray Regime}
In the soft-intermediate state, the evolution in X-ray and OIR regimes was quite typical. Smoothly decaying disk temperature and disk flux may imply presence of a steadily cooling accretion disk. Variable power law flux with a variable soft photon index can be thought as arising from a variable corona which have some overlap with the accretion disk that provides strong cooling. The emission mechanism that produces the power law component is most probably the thermal Comptonization \citep{Sunyaev80}. 

After the start of state transition, the disk temperature and the disk flux rapidly decayed and remained out of the PCA energy range in about ten days. As the disk is thought to be the source of seed photons in the Comptonization process, this is most often interpreted as the reduction in the number of soft seed photons. This results in a less cooling of the corona, and higher electron temperature which is consistent with the observed hardening of photon index. This simple Comptonization scenario could change or be modified with the onset of compact jet. Because the jet can either provide additional source of seed photons from non-thermal synchrotron emission at the base of jet (corona) and channel some portion of the accretion energy into its own power \citep{Fender03}. These processes could provide a sharp increase in the rms amplitude of variability \citep{Hynes03var,Russell11} and/or a softening of the X-ray spectrum \citep{Dincer08,Russell10}, however no such sharp changes in the X-ray spectral and temporal properties were observed during rebrightening in the OIR.

\subsection{Origin Of OIR Emission On The Initial Decay}
At the initial decay (between MJD 55,585 and 55,605, see Figure~\ref{mwlc}), the OIR flux was most probably dominated by the thermal emission originating from the outer parts of the accretion disk as the secondary star is not expected to contribute to the OIR flux significantly \citep{Shahbaz01}. This is supported by the higher spectral slopes observed compared to the slopes during the rebrightening (see Figure~\ref{SEDall}). Additionally, the spectral slope of 1.34 $\pm$ 0.16 in this stage is lower than the expected spectral index from the Rayleigh Jeans tail of the black body ($F_{\nu}\propto\nu^{2}$) and seems consistent with the X-ray irradiation of the disk \citep{Hynes05, Coriat09, Buxton12}. The strong variability in the light curve in the intermediate state also supports X-ray irradiation. Furthermore, recent optical and near-infrared spectroscopic observations at the soft state suggests that the irradiation might be enhanced due to a larger illumination area provided by a warped disk, and/or the winds launched at large radii of the accretion disk that up-scatter part of the X-ray and/or UV emission \citep{Rah12}.

\subsection{Rebrightening Due To A Jet?}
The occurrence of OIR rebrightening in the hard state decay is similar to those observed in other black hole binaries 4U 1543-47 \citep{Buxton04, Kalemci05} and XTE J1550-564 \citep{Jain01, Russell10}, XTE J1752-223 \citep{Russell11}. Such a rebrightening has also been observed in 2003, 2005 and 2007 hard state decays of GX 339-4 \citep{Coriat09, Buxton12}. In all these cases, the rebrigtening occurs at an X-ray luminosity range of 0.08 \% to 2\% $L_{Edd}$ (Kalemci et al. in prep.). 4U 1543-47 \citep{Park04}, XTE J1550-564 \citep{Corbel01} and XTE J1752-223 \citep{Miller11} have shown radio revival sometime during the rebrightening, but the coverage for these cases were not adequate to describe radio behavior during the OIR rise. For 4U 1543-47, XTE J1550-564 and XTE J1752-223, the SED of the excess during the rebrightening has indicated clear negative slope which is consistent with the optically thin synchrotron emission from a compact jet, and an extrapolation from radio to OIR matches well with a flat or slightly inverted spectral slopes. GX 339-4 during the rebrightening in the 2005 hard state decay showed similar radio-OIR SEDs \citep{Coriat09}. Note that rebrightening in the hard state decay of GX 339-4 in 2011 is also associated with radio revival (S. Corbel, private comm.), hence it is natural to assume that the rebrightening has jet origin.

\subsection{Understanding The Broadband SEDs}
We have produced two broadband SEDs from the rise and the peak of the rebrightening. Unlike 4U 1543-47 and XTE J1550-564 for which a clear negative slope is present, our baseline subtracted SEDs indicate a flat to slightly inverted spectrum (see Figure~\ref{seds}). Assuming that the excess is from the jet, we tried to obtain the break frequency for the change from an optically thick to an optically thin synchrotron emission, which is an important parameter in determining the base radius of the jet, magnetic field and minimum total jet power \citep{Fender06}. 

The flat SEDs may indicate that all OIR points lie on the optically thick part. In this case the break is around the V band, or at a higher frequency, which places a constraint on the slope of the optically thin synchrotron emission. A power law fit from V-band that passes through the softest X-ray band would give the spectral slopes of $\alpha=0.72 \pm 0.03$ and $\alpha=0.95 \pm 0.03$ for MJD 55,609.84 and 55,617.53, respectively (the dot-dashed lines in Figure~\ref{seds}). These slopes are a lower limit to the possible absolute slopes that do not provide X-ray fluxes higher than the observed ones. The former is a typical spectral slope expected from an optically thin synchrotron jet emission, but the latter is above the given range $0.6\lesssim \alpha \lesssim0.8$ \citep{Russell11} depending on the lepton energy distribution. If the spectral slopes as steep as 1.5 for XTE J1550-564 \citep{Russell10} and XTE J1752-223 \citep{Russell11} are possible at the beginning of the OIR rebrightening, then the particle distributions inferred from SEDs would not be problematic.

If the break is at the V band, however, the ratio of the minimum integrated synchrotron power to the X-ray luminosity is $L_{\rm j}/L_{\rm 1-200~keV}$=0.18 and 0.95 for the SEDs on MJD 55,609.84 and MJD 55,617.53, respectively. For this calculation, we considered only the optically thick part of the SEDs and assummed a conservative jet radiative efficiency of 0.05 following \citealt{Corbel02}. The reported ratios for GX 339-4 are in between 0.05-0.10 \citep{Corbel02} and the maximum value reported so far is for XTE J1118+480 which is 0.20 \citep{Fender01}. The ratio of 0.95 is too high to be accounted by scale-invariant jet models \citep{Markoff05}. The ratio decreases down to 0.32 for a break at the H band for this observation, and 0.05 for the observation on MJD 55,609.84. The typical break frequencies reported so far for GX 339-4 are in the mid-infrared \citep{Gandhi11} to infrared range \citep{Corbel02, Coriat09, Homan05, Buxton12}.

On the other hand, if we place the break at H-band or at a lower frequency, we encounter a problem with a simple post-shock emission model. In Figure~\ref{seds}, the dashed lines show the limiting cases of the power laws that can be fitted from H-band to softest X-ray band. The other three OIR bands remain significantly above the power law even when the errors are considered. This shows that the entire excess SED cannot be explained by optically thin synchrotron emission from a jet. \cite{Coriat09} have studied SEDs from a similar stage of the hard state decay and claimed that significant reprocessing may be present during the OIR rebrightening. In this work, we show that even taking out the baseline emission that comes with reprocessing, there is still an excess. If a spectral break occurs at/below the H-band, it either must have two components (such as pre-shock synchrotron, \citealt{Markoff03,Homan05}) or a simple flat to inverted spectrum breaking in the near or mid infrared is not a good description of the jet behavior during its launch. Such an additional component has recently been observed by \cite{Rah12}.  A full SED fitting (including radio fluxes) that takes into account the emission from different parts of the jet is required to find the break frequency, and this is beyond the scope of this paper.

We also extrapolated a power law (dotted lines) with spectral index obtained from X-ray spectral analysis down to the OIR band. In both days, the power law remained substantially below the OIR. This shows that the X-ray spectra cannot be explained by pure synchrotron emission without assuming a second emission component from the jet, at least during the early part of the OIR rebrightening. X-rays dominated by direct jet emission scenario such as stated in \cite{Maitra09} is also dismissed in \cite{Buxton12}. The synchrotron-self Compton models are also able to explain the broadband nature of jet emission \citep{Markoff05, Coriat09}. However a detailed analysis of such models are beyond the scope of this paper.  

\subsection{On The Modulations Of the OIR Light Curves}
We have found that the OIR light curves on the rise of the rebrightening are modulated with the binary period of the system ($\sim$1.77 days). This can be explained with the X-ray irradiation of the secondary star by the hard photons originating from the corona. Then, we need to explain why we do not detect variations with the binary period before the rebrightening, and after the peak of the OIR light curve.
Between MJD 55,580 and 55,605 (before the rebrightening), the power-law flux is relatively high, however, the spectral indices are still high. During this time, the size of the corona may be relatively small, and therefore cannot illuminate the secondary star effectively while still effectively illuminating the warped disk. During this time, variations in the OIR flux from the disk is larger than the variations with the binary period. When the corona becomes larger, as indicated by the changes in the spectral index and the flux, it not only illuminates the secondary star effectively, but also allows the magnetic flux to travel in close to the black hole, aiding in the launch of the jet \citep{Meier01,Beckwith09}. At the peak, the lack of modulations with the binary period is most likely due to the strongly variable jet synchrotron \citep{Rah12} dominating the emission in all bands. After the peak, the X-ray power law flux decreases so much that it cannot produce significant irradiation on the surface of the secondary star. In Figure~\ref{mwlc}, the decrease in the X-ray power law flux after MJD 55,607 is accompanied by a decrease in the amplitude of variations in V-band which is in favour of X-ray irradiation of secondary star.

\acknowledgements
TD thanks all scientists who contributed to the T\"{u}bingen Timing Tools and acknowledges T\"{U}B\.{I}TAK grant 111T222. We also thank Tomaso Belloni and the anonymous referee for their valuable comments on the manuscript. EK and SC acknowledge support from FP7 Initial Training Network Black Hole Universe, ITN 215212.

{\it Facilities:} \facility{RXTE (PCA)}, \facility{Swift (XRT)}, \facility{SMARTS}

\bibliographystyle{apj}

\end{document}